\pdfoutput=1

\documentclass[11pt]{article}

\usepackage[]{ACL2023}

\usepackage{times}
\usepackage{latexsym}
\usepackage{tabularx}

\usepackage[T1]{fontenc}

\usepackage[utf8]{inputenc}

\usepackage{microtype}

\usepackage{inconsolata}

\usepackage{graphicx,booktabs,multirow,amsmath,bm,amssymb,dsfont}
\usepackage{pifont}
\usepackage{comment}
\usepackage{tikz}
\newcommand*\circled[1]{\tikz[baseline=(char.base)]{
		\node[shape=circle,draw,inner sep=0.5pt] (char) {#1};}}

\usepackage{enumitem}

\title{CosyVoice: A Scalable Multilingual Zero-shot Text-to-speech Synthesizer based on Supervised Semantic Tokens}

\author{Zhihao Du, Qian Chen, Shiliang Zhang, Kai Hu, Heng Lu, Yexin Yang\\ {\bf Hangrui Hu, Siqi Zheng, Yue Gu, Ziyang Ma, Zhifu Gao, Zhijie Yan} \\
Speech Lab, Alibaba Group, China \\
\texttt{\{neo.dzh,sly.zsl,h.lu\}@alibaba-inc.com}}

\begin{document}
\maketitle
\begin{abstract}
Recent years have witnessed a trend that large language model (LLM) based text-to-speech (TTS) emerges into the mainstream due to their high naturalness and zero-shot capacity.
In this paradigm, speech signals are discretized into token sequences, which are modeled by an LLM with text as prompts and reconstructed by a token-based vocoder to waveforms.
Obviously, speech tokens play a critical role in LLM-based TTS models.
Current speech tokens are learned in an \textit{unsupervised} manner, which lacks explicit semantic information and alignment to the text.
In this paper, we propose to represent speech with \textit{supervised} semantic tokens, which are derived from a multilingual speech recognition model by inserting vector quantization into the encoder.
Based on the tokens, we further propose a \textbf{Co}dec-based \textbf{sy}nthesizer for \textbf{Voice} generation, CosyVoice\footnote{Models and codes are released at \url{https://github.com/FunAudioLLM/CosyVoice}. Demos can be found at \url{https://fun-audio-llm.github.io}}, which consists of an LLM for text-to-token generation and a conditional flow matching model for token-to-speech synthesis.
Experimental results show that \textit{supervised} semantic tokens significantly outperform existing \textit{unsupervised} tokens in terms of content consistency and speaker similarity for zero-shot voice cloning.
Moreover, we find that utilizing large-scale data further improves the synthesis performance, indicating the scalable capacity of CosyVoice. To the best of our knowledge, this is the first attempt to involve  \textit{supervised} speech tokens into TTS models.
\end{abstract}

\section{Introduction}

Text-to-Speech (TTS) technology has made remarkable strides in recent years, transitioning from robotic-sounding speech to producing voices that are nearly indistinguishable from human speakers. At the forefront of this advancement are Large Language Models (LLMs), which have been increasingly utilized in TTS systems to generate speech with a higher degree of naturalness and the ability to synthesize voices in a zero-shot fashion \citep{DBLP:journals/corr/abs-2305-07243,DBLP:journals/corr/abs-2301-02111,DBLP:journals/corr/abs-2402-08093}. These LLM-based TTS models function by converting speech signals into sequences of tokens, with the LLM utilizing text as a condition to model these token sequences. A token vocoder is then employed to reconstruct the raw waveforms from the tokenized speech \citep{DBLP:conf/nips/KongKB20, DBLP:journals/corr/abs-2210-13438}.

A critical aspect of the TTS process is the representation of speech tokens. Traditionally, tokens are acquired through unsupervised learning, which may not capture explicit semantic information or align well with corresponding text \citep{DBLP:journals/taslp/HsuBTLSM21,DBLP:journals/corr/abs-2210-13438}. Recognizing this gap, our work introduces supervised semantic tokens extracted from a multilingual speech recognition model, Whisper \citep{DBLP:conf/icml/RadfordKXBMS23}, by integrating vector quantization into the encoder. This innovation allows for more accurate semantic representation and alignment with text. Early studies have shown that quantizers with auxiliary automatic speech recognition (ASR) loss outperform k-means clustering on the universal speech model (USM) for speech-to-text translation and ASR tasks, as demonstrated in \citet{DBLP:journals/corr/abs-2306-12925}. Additionally, \citet{DBLP:journals/spl/YeGCLZ24} employed Gumbel-Softmax vector quantization to extract discrete speech representations that prioritize ASR-relevant information for ASR tasks. However, the impact of these approaches on text-to-speech (TTS) remains unclear.

Furthermore, leveraging these supervised tokens, we propose CosyVoice, a scalable and efficient zero-shot TTS synthesizer. CosyVoice is comprised of an LLM for converting text into semantic token sequences and a conditional flow matching model for the subsequent synthesis of speech from these tokens. 
In contrast to prior systems like TorToise TTS \citep{DBLP:journals/corr/abs-2305-07243}, which employs an LLM in conjunction with a denoising diffusion probabilistic models (DDPM)~\citep{DBLP:conf/nips/HoJA20}, CosyVoice utilizes a conditional flow matching approach, as it has been demonstrated to accelerate both training and inference compared to traditional diffusion models \citep{le2024voicebox}. While existing methods incorporate flow matching in TTS \citep{le2024voicebox, DBLP:journals/corr/abs-2309-05027,DBLP:journals/corr/abs-2309-03199,DBLP:journals/corr/abs-2309-17056}, they often rely on phoneme duration prediction, necessitating the use of supplementary phonemizers and forced aligners. CosyVoice, however, bypasses these dependencies, offering a more direct and efficient pathway from text to speech.

Our research contributes to the field of speech generation in several novel ways:
\begin{itemize}[leftmargin=*,noitemsep]
\item We are the first to integrate supervised speech tokens into TTS models, enhancing content consistency and speaker similarity in zero-shot voice cloning.
\item We propose CosyVoice, a scalable zero-shot TTS synthesis system that combines an LLM for text-to-token generation with a conditional flow matching model for token-to-speech synthesis, forsaking the need for additional phonemizers and forced aligners.
\item To further refine the quality of generated speech, we incorporate the x-vector \citep{DBLP:conf/icassp/SnyderGSPK18} into the LLM to separate the modeling of speech into semantic, speaker, and prosody components. The LLM models the semantic content and prosody, while the conditional flow matching model captures timbre and environmental information. We optimize the flow matching process with techniques such as classifier-free guidance \citep{DBLP:journals/corr/abs-2207-12598}, a cosine scheduler, and masked conditions.
\end{itemize}

Our experimental results demonstrate the superiority of supervised semantic tokens over unsupervised counterparts. Additionally, the scalability of CosyVoice is evidenced by improved synthesis performance when utilizing large-scale data. This work, therefore, represents a significant step forward in the development of natural-sounding, versatile TTS systems.


\begin{figure*}[t!]
  \centering
  \includegraphics[width=\linewidth]{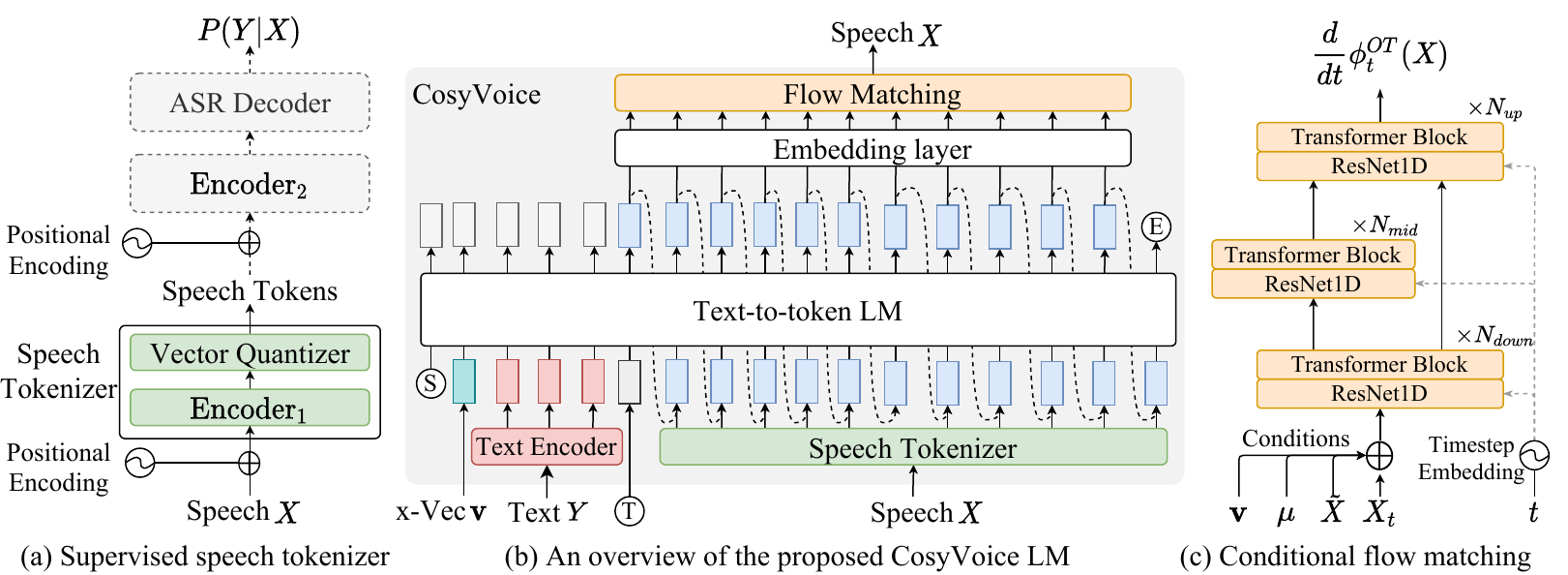}
  \caption{An overview of the proposed CosyVoice model. (a) demonstrates the $\mathcal{S}^3$ tokenizer, where dashed modules are only used at the training stage. (b) is a schematic diagram of CosyVoice, consisting of a text-to-token LLM and a token-to-speech flow matching model. \circled{S}, \circled{E} and \circled{T} denote the ``start of sequence'', ``end of sequence'' and ``turn of speech'' tokens. Dashed lines indicate the autoregressive decoding at the inference stage. (c) provides an enlarged view of our flow matching model conditioning on a speaker embedding $\mathbf{v}$, semantic tokens $\mu$, masked speech features $\tilde{X}$ and intermediate state $X_t$ at timestep $t$ on the probabilistic density path.}
  \label{fig:overall}
\end{figure*}

\section{CosyVoice: A Scalable TTS model using Supervised Semantic Tokens}
As shown in Figure \ref{fig:overall}(b), our CosyVoice consists of four components, namely text encoder, speech tokenizer, large language model and conditional flow matching model.
Specifically, the text encoder is used to align the semantic spaces of text and speech tokens, while the speech tokenizer is utilized to extract semantic tokens as illustrated in Figure \ref{fig:overall}(a).
We employ a large language model to learn the whole sequence of text encodings and speech tokens, reformulating TTS as an auto-regressive sequence generation problem given text as prompts.
Then, as shown in Figure \ref{fig:overall}(c), a conditional flow matching model is utilized to convert speech tokens into a Mel spectrogram via a denoising process on the optimal path.
To obtain a perceptible signal, the HifiGAN vocoder \citep{DBLP:conf/nips/KongKB20} is used to synthesize a waveform with the generated Mel spectrogram as input.

\subsection{Supervised Semantic Tokens for Speech}
\label{sec:sst}
In CosyVoice, a supervised automatic speech recognition (ASR) model is employed to derive the supervised semantic speech ($\mathcal{S}^3$) tokenizer for speech. The model is a finetuned version of our proprietary SenseVoice ASR model. It is trained on multilingual audio data and possesses rich audio content understanding capabilities.
Different from the original ASR model, we split the encoder into two parts and insert a vector quantization layer between them. Given a Mel spectrogram $X$ as input, it undergoes the positional encoding and $\mathrm{Encoder}_1$ to obtain a context-aware representations $H$:
\begin{equation}
	H = \mathrm{Encoder_1}\left(\mathrm{PosEnc}(X)\right)
\end{equation}
Then, a vector quantizer (VQ) is involved to obtain discrete tokens.
For the hidden representation $\mathbf{h}_l$ at the frame $l$, the index of nearest embedding in the codebook $C$ is treated as the speech token $\mu_l$ at this timestep:
\begin{equation}
	\mu_l = \mathrm{VQ}(\mathbf{h}_l, C)=\mathrm{arg}\min_{\mathbf{c}_n\in C}{|| \mathbf{h}_l - \mathbf{c}_n ||_2}
\end{equation}
where $||\cdot||_2$ denotes the L2 norm. At the training stage, codebook embeddings are updated via exponentially moving average (EMA):
\begin{equation}
	\mathbf{c}_{\mu_l} := \alpha \mathbf{c}_{\mu_l} + (1-\alpha) \mathbf{h}_l
\end{equation}
where $\alpha$ is a pre-defined decay coefficient. The corresponding codebook embeddings of speech tokens are used as the quantized hidden representations $\bar{H}=\{\mathbf{c}_{\mu_1}, \mathbf{c}_{\mu_2}, \dots, \mathbf{c}_{\mu_L}\}$ and passed through the remaining encoder layers $\mathrm{Encoder}_2$:
\begin{equation}
	\tilde{H} = \mathrm{Encoder_2}\left(\mathrm{PosEnc}(\bar{H})\right)
\end{equation}
Note that, before the remaining encoder layers, we add an extra positional encoding to enhance the temporal information.
After $\mathrm{Encoder_2}$, a transformer-based ASR decoder is followed, predicting the posterior probability of text labels:
\begin{equation}
	P(Y|X)=\mathrm{{ASRDecoder}}\left(\tilde{H},Y^{Z-1}\right)
\end{equation}
where $Y^{Z-1}$ represents the left-shifted text labels in the teacher-forcing training scheme.

\subsection{Large Language Model for TTS}
In this section, we formulate the TTS task as an auto-regressive speech token generation problem with a large language model (LLM). 
For LLM, the sequence construction is the most important matter, which is constructed as follows:
\begin{equation}
	\left[\circled{S}, \mathbf{v}, \{\bar{\mathbf{y}}_u\}_{u\in[1:U]}, \circled{T}, \{\mu_l\}_{l\in[1:L]}, \circled{E} \right]
\end{equation}
\circled{S} and \circled{E} denote the start and end of sequence, respectively. $\mathbf{v}$ is a speaker embedding vector extracted from the speech $X$ with a pre-trained voice-print model\footnote{Available at https://github.com/alibaba-damo-academy/ 3D-Speaker/tree/main/egs/3dspeaker/sv-cam++}. 
The text encodings $\bar{Y}=\{\bar{\mathbf{y}}_u\}_{u\in[1:U]}$ is obtained by passing the text through a Byte Pair Encoded (BPE) tokenizer and text encoder:
\begin{equation}
	\bar{Y} = \mathrm{TextEncoder}(\mathrm{BPE}(Y))
\end{equation}
Since text and speech tokens lie at different semantic levels, the text encoder is used to align their semantic spaces and benefit the LLM modeling.
A start identifier \circled{T} is inserted between text encodings and speech tokens $\{\mu_l\}_{l\in[1:L]}$ that is extracted with the supervised semantic tokenizer as described in \ref{sec:sst}. At the training stage, we employ the teacher-forcing scheme, in which the left-shifted sequence is employed as the mode inputs and the original sequence serves as the expected outputs. 
Note that only the cross entropy losses of speech tokens and \circled{E} are considered during the training:
\begin{equation}
	\mathcal{L}_{LM} = -\frac{1}{L+1}\sum_{l=1}^{L+1}{\log{q(\mu_l)}}
\end{equation}
where $\mu_{L+1}$ is the ``end of sequence'' token \circled{E}. $q(\mu_l)$ denotes the posterior probability of $\mu_l$, which is predicted by the softmax layer following LLM.

\subsection{Optimal-transport Conditional Flow Matching}
In CosyVoice, an optimal-transport conditional flow matching model (OT-CFM) is employed to learn the distribution of Mel spectrogram and generate samples from it with generated speech tokens as conditions.
OT-CFM can achieve better performance compared to diffusion probabilistic models (DPMs) with simpler gradients, easier training and faster generation \cite{DBLP:conf/iclr/LipmanCBNL23,tong2023improving,DBLP:journals/corr/abs-2309-03199}.
In continuous-time normalizing flows (CNFs), a probability density path is constructed from a prior distribution $p_0(X)$ to the data distribution of Mel spectrogram $q(X)$.
The probability density path is defined by a time-dependent vector field $\nu_t(X): [0,1]\times \mathbb{R}^{L*D}\rightarrow \mathbb{R}^{L*D}$, which generates the flow $\phi_t$ through the following ordinary differential equation (ODE):
\begin{equation}
\label{eq:prob-path}
\begin{aligned}
\frac{d}{dt}{\phi_t{(X)}} &= \nu_t(\phi_t(X), t) \\
\phi_0(X)&\sim p_0(X)=\mathcal{N}(X;0,I) \\
\phi_1(X)&\sim p_1(X)
\end{aligned}
\end{equation}
where $t\in[0, 1]$. By solving the initial value problem Eq. (\ref{eq:prob-path}), we can approximate the speech distribution $q(X)$ with $p_1(X)$ and sample from it.
To learn the vector field $\nu_t(X)$, we define the optimal-transport (OT) flow and force a neural network to match it by minimizing the following loss:
\begin{equation}
\begin{aligned}
	&\mathcal{L}_{OT-CFM} \\
	&= \mathbb{E}_{t,p_0(X_0),q(X_1)}| \omega_t(\phi^{OT}_t(X_0,X_1)|X_1) \\
	&- \nu_t(\phi^{OT}_t(X_0,X_1)|\theta) |
\end{aligned}
\end{equation}
where
\begin{equation}
\begin{aligned}
	&\phi^{OT}_t(X_0,X_1)=(1-(1-\sigma)t)X_0+tX_1 \\
	                &\omega_t(\phi^{OT}_t(X_0,X_1)|X_1)=X_1-(1-\sigma)X_0
\end{aligned}
\end{equation}
The speaker embedding $\mathbf{v}$, speech tokens $\{\mu_l\}_{1:L}$, and masked Mel spectrogram $\tilde{X_1}$ are also fed into the neural network to match the vector field with learnable parameters $\theta$:
\begin{equation}
\begin{aligned}
	&\nu_t(\phi^{OT}_t(X_0,X_1)|\theta) \\
	&= \mathrm{NN}_\theta\left(\phi^{OT}_t(X_0,X_1),t;\mathbf{v},\{\mu_l\}_{1:L},\tilde{X_1}\right)
\end{aligned}
\end{equation}
$\tilde{X_1}$ is a masked version of $X_1$ by setting continuous frames to zeros from a random start point to the end.
Considering the generation process at the beginning is harder than follows, we involve a cosine scheduler for the timestep $t$:
\begin{equation}
t:=1-\cos\left(\frac{1}{2}t\pi\right)
\end{equation}
Under the scheduled flow, there are more generation steps at the beginning.

Classifier-free guidance (CFG) has been proven to improve the generation quality of diffusion probabilistic models \cite{ho2022classifier,nichol2021improved,le2024voicebox}.
Therefore, we propose to adapt the CFG into conditional flow matching models.
At the training stage, we randomly drop the conditions $\Psi=\{\mathbf{v}, \{\mu_l\}_{1:L}, \tilde{X_1}\}$ with a fixed probability of $0.2$. In this manner, we can learn both conditional and unconditional flows. During generation, the vector field is modified as follows:
\begin{equation}
\begin{aligned}
	&\tilde{\nu}_t(\phi^{OT}_t(X_0,X_1)|\theta;\Psi)\\
	&=(1+\beta)\cdot\nu_t(\phi^{OT}_t(X_0,X_1)|\theta;\Psi)\\
	&-\beta \cdot \nu_t(\phi^{OT}_t(X_0,X_1)|\theta)
\end{aligned}
\end{equation}
where $\beta$ is the guidance strength of $0.7$.

\subsubsection{Zero-shot In-context Learning}
\begin{figure}[t]
	\centering
	\includegraphics[width=\linewidth]{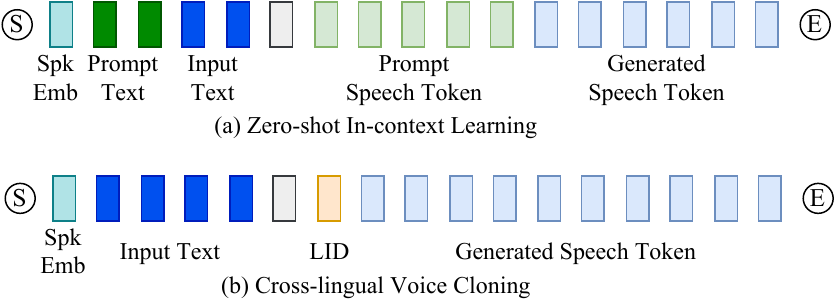}
	\caption{Sequence construction for (a) zero-shot in-context learning and (b) cross-lingual voice cloning. LID represents language identifier.}
	\label{fig:icl-seq}
\end{figure}

CosyVoice models exhibit zero-shot in-context learning capabilities, allowing for the replication of an arbitrary voice with only a brief reference speech sample. This process entails the careful construction of input sequences for the token language model (LM), depicted in Figure \ref{fig:icl-seq}.
For prompt speech and input text in the same language, we merge them to form a unified input, treating the prompt speech tokens as pre-generated. With this input sequence, the autoregressive LM iteratively predicts subsequent tokens until it encounters the ``end of sequence'' token $\circled{E}$.
However, when the prompt speech and input text differ linguistically, we omit the text and tokens associated with the prompt to prevent prosodic characteristics of the original language from influencing the target language.
It is important to note that the prompt text, which corresponds to the prompt speech's content, can be transcribed either through human annotation or ASR models, such as SenseVoice. Similar to the prompt text, the prompt tokens are extracted from the prompt speech with $\mathcal{S}^3$ tokenizer.

After generating the speech tokens, they are appended after the prompt tokens, forming a composite condition for the flow-matching model. Additionally, the speaker embedding and the Mel spectrogram of the prompt speech are incorporated to further enhance timbre and environmental consistency.

\begin{table*}[t]
	\footnotesize
	\centering
	\scalebox{1.0}{
		\begin{tabularx}{\textwidth}{X}
			\toprule
			\small{\textbf{Speaker Identity}}  \\
			1. Selene 'Moonshade', is a \textbf{mysterious}, \textbf{elegant dancer} with a connection to the night. Her movements are both \textbf{mesmerizing} and \textbf{deadly}.$<$endofprompt$>$Hope is a good thing.\\
			2. Theo 'Crimson', is a \textbf{fiery}, \textbf{passionate} rebel leader. Fights with fervor for justice, but struggles with \textbf{impulsiveness}.$<$endofprompt$>$You don't know about real loss. \\
			\midrule
			\small{\textbf{Speaking Style}}  \\
			1. A \textbf{happy} \textbf{girl} with \textbf{high tone} and \textbf{quick speech}.$<$endofprompt$>$The sun is shining brightly today. \\
			2. A \textbf{sad woman} with \textbf{normal tone} and \textbf{slow speaking speed}.$<$endofprompt$>$I failed my important exam. \\
			\midrule
			\small{\textbf{Fine-grained Paralinguistics}} \\
			1. Well that's kind of scary \textbf{[laughter]}. \\
			2. I don't think I over eat yeah \textbf{[breath]} and um I do exercise regularly. \\
			3. Well that pretty much covers \textbf{$<$laughter$>$the subject$<$/laughter$>$} well thanks for calling me. \\	
			4. The team's \textbf{$<$strong$>$unity$<$/strong$>$} and \textbf{$<$strong$>$resilience$<$/strong$>$} helped them win the championship. \\
			\bottomrule
		\end{tabularx}
	}
	\caption{Examples of speaker identity, speaking style, and fine-grained paralinguistics. }
	\label{tab:example_instruct}
\end{table*}

\subsection{Rich Generation with Instruction}
To enable further controllability on CosyVoice, we experiment with integrating additional instruction fine-tuning \citep{DBLP:journals/corr/abs-2308-14430}. 
CosyVoice-instruct extends CosyVoice-base with enhanced instruction-following capabilities. Specifically, it supports controllability over various aspects such as speaker identity (i.e., speaker's characteristics), speaking style (including emotion, gender, speaking rate, and pitch), and fine-grained paralinguistic features. These features include the ability to insert laughter, breaths, speaking while laughing, and emphasize certain words.

We fine-tuned CosyVoice-base using this training data without incorporating speaker embedding in the autoregressive language model. Table \ref{tab:example_instruct} shows some examples of speaker identity, speaking style, and fine-grained paralinguistic features.

\section{Dataset}
\subsection{Small-scale Single-lingual Dataset}
We conduct experiments on the LibriTTS \citep{zen2019libritts} corpus, which contains 585 hours from 2,456 English speakers. We follow the official data partition, where ``train-clean-100'', ``train-clean-360'' and ``train-other-500'' are merged for training and the ``dev-clean'' is used for model selections. ``test-clean'' is used to construct the evaluation set as described in \citep{DBLP:conf/aaai/DuGSLLCWZ024}.

\subsection{Large-scale Multi-lingual Dataset}
\begin{table}[t]
	\centering
	\setlength\tabcolsep{10pt}
	\scalebox{0.9}{
		\begin{tabular}{lr}
			\toprule
			Language & Duration (hr) \\
			\midrule
			ZH & 130,000 \\
			EN & 30,000 \\
			Yue & 5,000 \\
			JP & 4,600 \\
			KO & 2,200 \\
			\bottomrule
	\end{tabular}}
	\caption{Hours of CosyVoice training data across languages in the large-scale experiments.}
	\label{tab:dataset}
\end{table}

\begin{table}[t]
	\centering
	\setlength\tabcolsep{8pt}
	\scalebox{0.9}{
		\begin{tabular}{lr}
			\toprule
			Type & Duration (hr) \\
			\midrule
			Speaker Identity & 101 \\
			Speaking Style & 407 \\
			Fine-grained Paralinguistics & 48 \\
			\bottomrule
	\end{tabular}}
	\caption{Duration statistics of instruction training data by type.}
	\label{tab:dataset_instruct}
\end{table}
To train the CosyVoice models, we have amassed a considerable dataset comprising multiple languages. Throughout the collection process, we utilize specialized in-house tools for speech detection, signal-to-noise ratio (SNR) estimation, speaker diarization, and separation. Subsequently, pseudo text labels are generated using SenseVoice-Large and Paraformer. These labels undergo a refinement process with the aid of force-alignment (FA) models, which helps eliminate low-quality data and enhances the accuracy of punctuation. A comprehensive breakdown of the training data's duration across various languages is presented in Table \ref{tab:dataset}. Table \ref{tab:dataset_instruct} presents the duration of the training data for different types of instructions. 

\section{Experimental Settings}
\subsection{Supervised Semantic Speech Tokenizer}
For the small-scale single-lingual dataset, we employ the ESPNet Conformer ASR model as the backbone and insert the vector quantizer after the first six encoder layers. There is a single codebook with 4,096 codes. The first six encoder layers and vector quantizer are employed as the speech tokenizer. As for the text tokenizer, a word sentence-piece model is trained on the text of training, which has a vocabulary size of 4,000. We train the quantizer-augmented ASR model on the Librispeech \cite{panayotov2015librispeech} corpus for 50 epochs from scratch.

For the large-scale multi-lingual dataset, we employ the SenseVoice-Large rich recognition model \citep{funaduiollm} as the backbone. Similar to small-scale dataset, we still insert the vector quantizer after the first six encoder layers with a single codebook of 4,096 codes. More hyper-parameter selections, such as quantizer-inserted layer and the number of codes, are left for future work. Different from single-lingual experiments, we use the pre-trained checkpoint to initialize the SenseVoice-Large model rather than train it from scratch. After inserting the quantizer, we further fine-tune the whole parameters for 210,000 training steps on eight A800 GPUs.

\begin{table}[t]
	\centering
	\setlength\tabcolsep{10pt}
	\scalebox{0.9}{
		\begin{tabular}{l|cc}
			\toprule
			Settings & Tiny & Normal \\
			\midrule
			\multicolumn{3}{c}{Text Encoder} \\
			\hline
			Layers & 6 & 6 \\
			Attention Dim. & 512 & 1,024 \\
			Attention Heads & 8 & 16 \\
			Linear Units & 2,048 & 4,096 \\
			\hline
			\multicolumn{3}{c}{Language Model} \\
			\hline
			Layers & 12 & 14 \\
			Attention Dim. & 512 & 1,024 \\
			Attention Heads & 8 & 16 \\
			Linear Units & 2,048 & 4,096 \\
			\bottomrule
	\end{tabular}}
	\caption{Details of model architecture settings in the tiny and normal CosyVoice models.}
	\label{tab:model}
\end{table}
\subsection{CosyVoice Model Settings}
We train the tiny and normal size models in single-lingual and multi-lingual experiments. Details of model architecture settings are shown in Table \ref{tab:model}. The tiny model is trained on LibriTTS training set for 50 epochs with four V100-32M GPUs, while the multi-lingual model is trained on our internal dataset for 800,000 steps with 64 V100-32M GPUs.
Tiny and normal models are trained with the learning rate of $10^{-3}$ and $10^{-4}$, respectively. The warmup step is set to 10,000. 

\section{Experimental Results}
\subsection{Evaluation on $S^3$ Tokenizer}

\begin{table}[t]
	\centering
	\setlength\tabcolsep{3pt}
	\scalebox{0.9}{
		\begin{tabular}{l|ccc}
			\toprule
			Model & dev\_clean & test\_clean & test\_other \\
			\midrule
			Conformer & 2.62 & 2.89 & 6.57 \\
			Conformer-VQ & 3.13 & 3.18 & 7.56 \\
			\bottomrule
	\end{tabular}}
	\caption{Impact of inserting vector quantization on speech recognition in terms of word error rate (\%).}
	\label{tab:en-token}
\end{table}

In table \ref{tab:en-token}, we demonstrate how the vector quantization affects the recognition performance on LibriTTS test sets. From the table, we can see that inserting a vector quantizer into the ASR encoder only affects the recognition performance slightly. As a result, the VQ-inserted Conformer ASR model achieves comparable WERs of 3.18\% and 7.56\% on ``test-clean'' and ``test-other'' sets, respectively. This indicates that tokenizers trained in a supervised manner can maintain sufficient semantic information and the alignment to text.

\begin{table}[t]
	\centering
	\setlength\tabcolsep{5pt}
	\scalebox{0.8}{
		\begin{tabular}{l|cc|cc|cc}
			\toprule
			& \multicolumn{2}{c|}{Whisper-L-V3} & \multicolumn{2}{c|}{SenseVoice-L} & \multicolumn{2}{c}{$\mathcal{S}^3$ tokens} \\
			Test set & w/o lid & w/ lid & w/o lid & w/ lid & w/o lid & w/ lid \\\hline
			zh-CN & 12.82 & 12.55 & 8.76 & 8.68 & 12.24 & 12.06 \\
			en & 13.55 & 9.39 & 9.79 & 9.77 & 15.43 & 15.38 \\
			\bottomrule
	\end{tabular}}
	\caption{The evaluation on $\mathcal{S}^3$ tokens' capability to preserve semantic information. We employ word and character error rates for zh-CN and en languages on the Common Voice benchmarks.}
	\label{tab:tokenizer-performance}
\end{table}

To assess the multi-lingual  $\mathcal{S}^3$ tokenizer's ability to preserve semantic information, we compared the recognition performance of the quantizer-augmented SenseVoice-L against its original version and the Whisper-Large V3 model. 
The models underwent evaluation using the Common Voice zh-CN and en benchmarks, with the findings detailed in Table \ref{tab:tokenizer-performance}.
From the table, we can see that our $\mathcal{S}^3$ tokens demonstrate robust recognition performance in both the Chinese and English test sets. Notably, on the common\_voice\_zh-CN set, $\mathcal{S}^3$ tokens surpass the performance of the Whisper-Large V3 model \cite{funaduiollm}, achieving a 4.14\% relative reduction in error rate. This suggests a substantial correlation between $\mathcal{S}^3$ tokens and semantic content.
It is worth noting that there is only a single codebook in the $\mathcal{S}^3$ tokenizer with a dictionary size of 4,096 entries.

\subsection{Comparison with Baselines}
\begin{table*}[t]
	\centering
	\setlength\tabcolsep{5pt}
	\scalebox{0.9}{
		\begin{tabular}{l|cccccc}
			\toprule
			\bf{Model} & \bf{Text token} & \bf{Speech Token} & \bf{WER (\%)} & \bf{\#INS+DEL} & \bf{\#SUB} & \bf{SS} \\
			\midrule
			Original & - & - & 3.01 & 66 & 200 & 69.67 \\
			VALL-E \citep{DBLP:journals/corr/abs-2301-02111} & Phone & Encodec & 18.70 & 342 & 1312 & 53.19 \\
			UniAudio \citep{DBLP:journals/corr/abs-2310-00704} & Phone & Encodec & 8.74 & 254 & 519 & 47.56 \\
			SpearTTS \citep{DBLP:journals/tacl/KharitonovVBMGP23} & Phone & Hubert & 6.14 & 133 & 410 & 51.71 \\
			\hline
			Exp-1-LibriTTS & Phone & Hubert & 7.41 & 325 & 409 & 67.85  \\
			Exp-2-LibriTTS & Phone & $S^3_{en}$ & 5.05 & 122 & 325 & 67.85 \\ 
			Exp-3-LibriTTS & BPE$_{en}$ & $S^3_{en}$ & 3.93 & 108 & 239 & 67.85  \\ 
			Exp-4-LibriTTS & BPE & $S^3$ & 4.76 & 134 & 287 & 65.94  \\ 
			Exp-4-Large-scale & BPE & $S^3$ & \bf{3.17} & \bf{96} & \bf{184} & \bf{69.49}  \\ 
			\bottomrule
	\end{tabular}}
	\caption{Comparison with other TTS models on the LibriTTS test-clean set in terms of content consistency and speaker similarity (SS). Non-autoregressive ASR model, Paraformer-en, is employed for fast evaluation.}
	\label{tab:compare}
\end{table*}
We compare the proposed CosyVoice models with other TTS systems on content consistency and speaker similarity. For content consistency, an ASR model is employed to recognize the generated utterances. We report the word error rate (WER), and the number of insertion, deletion and substation errors. As for the speaker similarity, we employ the ERes2Net model \citep{eres2net} to extract speaker embeddings of prompt and generated utterances, and their raw cosine similarity is treated as the speaker similarity. Experimental results are shown in Table \ref{tab:compare}.

Compared with other TTS models, the proposed CosyVoice framework achieves comparable content consistency and higher speaker similarity even using the same text and speech tokenizers. Comparing Exp-1, Exp-2 and Exp-3, we can see that both the text speech tokenizers are critical for content consistency and negligible for speaker similarity. In Exp 4 experiments, we replace the single-lingual text and speech tokenizers with the multi-lingual one. Only using the LibriTTS corpus to train the model degrades both the content consistency and speaker similarity. By involving the internal large-scale dataset, the performance is significantly improved, achieving the human parity quality.

\subsection{Evaluation on Generation Quality of CosyVoice}
We evaluate the quality of CosyVoice's speech synthesis by examining content consistency and speaker similarity.
The ``test-clean'' subset of LibriTTS \citep{zen2019libritts} and the test set of AISHELL-3 \citep{DBLP:conf/interspeech/ShiBXZL21} are employed to construct an evaluation set for English and Chinese, respectively. For each text in these sets, we randomly select a prompt speech. Content consistency was evaluated using Whisper-Large V3 \citep{DBLP:conf/icml/RadfordKXBMS23} for English and Paraformer~\citep{DBLP:conf/interspeech/GaoZ0Y22} for Chinese recognition. Speaker similarity was quantified by calculating the cosine similarity between speaker embeddings of the generated and prompt speeches, extracted using ERes2Net \citep{eres2net}.

Similar to other autoregressive language models, we employ a random sampling decoding strategy for our token LM and assessed the synthesis process using five different random seed values: 0, 7, 42, 123, and 1,337. The resultant evaluation metrics were averaged to determine the mean and standard deviation. Additionally, we conducted an ASR re-ranking to demonstrate potential performance improvements in offline mode.

Tables \ref{tab:res-libritts} and \ref{tab:res-aishell} present the results for English and Chinese, respectively. On the English dataset, CosyVoice attained human-level performance with similar content recognition and higher speaker similarity. ASR re-ranking notably enhanced content consistency, yielding a reduced word error rate (WER) of 1.51\%. CosyVoice outperformed ChatTTS in WER and the number of insertion and deletion errors, indicating superior content consistency. We did not assess speaker similarity for ChatTTS as it doesn't release voice cloning capabilities.

\begin{table}[t]
	\centering
	\setlength\tabcolsep{6pt}
	\scalebox{0.8}{
		\begin{tabular}{lccc}
			\toprule
			Model & WER (\%) & \#Ins.\&Del. & SS \\
			\midrule
			Original & 2.66 & 92 & 69.67 \\ 
			ChatTTS & 8.32 & 441 & - \\
			CosyVoice & 2.89$\pm$0.18 & 88.60$\pm$3.88 & 74.30$\pm$0.15 \\
			\ \ + 5$\times$ re-ranking & 1.51 & 47 & 74.30 \\
			\bottomrule
	\end{tabular}}
	\caption{The comparison of original and CosyVoice generated speeches on the LibriTTS test-clean set in terms of word error rate (WER) and speaker similarity (SS). ``$\pm$'' joins the mean and standard deviation for each evaluation metric. Whisper-Large V3 is employed as the ASR model.}
	\label{tab:res-libritts}
\end{table}

\begin{table}[t]
	\centering
	\setlength\tabcolsep{6pt}
	\scalebox{0.8}{
		\begin{tabular}{lccc}
			\toprule
			Model & CER (\%) & \#Ins.\&Del. & SS \\
			\midrule
			Original & 2.52 & 25 & 74.15 \\ 
			ChatTTS & 3.87 & 111 & - \\
			CosyVoice & 3.82$\pm$0.24 & 24.4$\pm$2.24 & 81.58$\pm$0.16 \\
			\ \ + 5$\times$ re-ranking & 1.84 & 11 & 81.58 \\
			\bottomrule
	\end{tabular}}
	\caption{The comparison of original and CosyVoice generated speeches on the AISHELL-3 test set in terms of character error rate (CER) and speaker similarity (SS). Paraformer-zh is employed as the ASR model.}
	\label{tab:res-aishell}
\end{table}

\begin{table*}[t!]
	\centering
	\scalebox{0.9}{
		\begin{tabular}{l c c c c c c}
			\toprule
			Model & Happy & Sad & Angry & Surprised & Fearful & Disgusted \\
			\midrule
			CosyVoice-base & 1.00$\pm$0.00 & 0.45$\pm$0.05 & 0.59$\pm$0.03 & 0.26$\pm$0.02 & 0.88$\pm$0.01 & 0.46$\pm$0.06  \\
			CosyVoice-instruct & 1.00$\pm$0.00 & 0.98$\pm$0.02 & 0.83$\pm$0.04 & 0.64$\pm$0.03 & 0.87$\pm$0.03 & 0.93$\pm$0.02  \\
			~~w/o instruction & 0.98$\pm$0.01 & 0.77$\pm$0.04 & 0.49$\pm$0.12 & 0.28$\pm$0.06 & 0.83$\pm$0.04 & 0.45$\pm$0.16 \\
			\bottomrule
	\end{tabular}}
	\caption{Comparison of emotion control accuracy between CosyVoice-base-300M and CosyVoice-instruct-300M. ``$\pm$'' joins the mean and standard deviation for each evaluation metric.}
	\label{tab:emo_acc}
\end{table*}

\begin{table*}[t!]
	\centering
	\setlength\tabcolsep{3pt}
	\scalebox{0.9}{
		\begin{tabular}{lcccc}
			\toprule
			Training Data & dev\_clean & dev\_other & test\_clean & test\_other \\
			\midrule
			Librispeech & 2.77 & 5.84 & 2.79 & 5.97 \\ 
			Syn on LS text & 2.79 & 6.37 & 3.00 & 6.59 \\ 
			Librispeech + Syn on LS text & 2.44 & 5.52 & 2.56 & 5.68 \\ 
			Librispeech + Syn on LS text $\times 2$ & 2.51 & 5.23 & 2.68 & 5.26 \\ 
			Librispeech + Syn on LS, MLS text & \textbf{1.93} & \textbf{4.43} & \textbf{2.04} & \textbf{4.53} \\
			\bottomrule
	\end{tabular}}
	\caption{Evaluation on CosyVoice generation quality by treating it as a data generator. Word error rates (\%) on the human-uttered test sets are employed as the evaluation metrics.}
	\label{tab:data-syn}
\end{table*}

As for the results in Chinese, the generated utterances of CosyVoice achieve a comparable CER as well as the errors of insertion and deletion compared with the original utterances. It seems that ChatTTS has a better generation ability on Chinese than English in terms of CER. Although ChatTTS and CosyVoice achieve a similar CER, ChatTTS produces more insertion and deletion errors, This is due to the problem of speaker leaking, where modal particles of another speaker is generated unexpectedly. On the contrary, CosyVoice doesn't suffer from this problem with much fewer insertion and deletion errors. With ASR re-ranking, CosyVoice reached a remarkably low CER of 1.84\%. As seen with English, CosyVoice also exhibited greater speaker similarity than the original utterances, showcasing its effective voice-cloning proficiency.

\subsection{Emotion Controllability of CosyVoice}
To verify the emotion controllability, we use the public speech emotion recognition model emo2vec\footnote{\url{https://modelscope.cn/models/iic/emotion2vec_base_finetuned}} \citep{DBLP:journals/corr/abs-2312-15185}. We generated and evaluated 100 English utterances for each of the six emotions: happy, angry, sad, surprised, fearful, and disgusted. The content of the synthesized text is designed to match the target emotion. We then measure the accuracy of the predicted emotions from the synthesized speech for each emotion. 

Table \ref{tab:emo_acc} shows the comparison of emotion control accuracy between CosyVoice-base and CosyVoice-instruct. For CosyVoice-instruct, the input consists of content text accompanied by a speaking style instruction (e.g., ``Happy.$<$endofprompt$>$Content Text''). In contrast, CosyVoice-base only receives the content text as input. 
The results indicate that CosyVoice-instruct with emotional instructions demonstrates a significant improvement over both CosyVoice-base and CosyVoice-instruct without emotional instructions.

\subsection{CosyVoice as a Data Generator}
A straightforward application of CosyVoice is as a data generator to augment the training data of other tasks, such as ASR, speech-to-speech translation (S2ST). Taking the ASR task an example, we conduct an experiment on the Librispeech corpus to evaluate CosyVoice's capability in generating high-quality data. The experimental results are shown in Table \ref{tab:data-syn}, where ``Librispeech'' denotes the original 960-hour data. ``Syn on LS text'' and ``Syn on LS text'' denote the generated data with the text from Librispeech and MLS training sets, respectively. From the table, we can see that only training on the synthesized data, the ASR model can achieve a comparable result than the original Librispeech training set. Upon integration of them, a notable enhancement in recognition accuracy is observed. An interesting finding is that involving the synthesized data on the MLS text significantly improves the recognition performance. This may indicates that the text diversity is more critical for ASR task than the duration of speech itself. This improvement can be attributed to the varied linguistic content introduced by CosyVoice synthesized samples. The findings from our evaluation underscore the high quality of the samples generated by CosyVoice.

\section{Conclusion}
In this paper, we introduce CosyVoice, a scalable multi-lingual speech generation model, which supports zero-shot in-context learning, cross-lingual voice cloning, instructed generation and fine-grained controlling of emotion, paralinguistic features. Experimental results show that the system architecture of CosyVoice is important for speaker similarity, while the text and speech tokenizers affect the content consistency much. Besides, we find that scaling up the model size and data volume can improve the performance significantly. As a result, CosyVoice achieves the human parity generation quality.

\bibliography{anthology,custom}
\bibliographystyle{acl_natbib}




\end{document}